\documentclass[12pt]{article}
\usepackage[utf8]{inputenc}
\usepackage{amsmath}
\usepackage{amsfonts}
\usepackage[title]{appendix}
\usepackage{amssymb}
\usepackage{graphicx}
\usepackage{setspace}
\usepackage{subcaption}
\usepackage[space]{grffile}
\usepackage[round]{natbib}
\usepackage{authblk}
\usepackage{color}
\usepackage{geometry} 
\title{Bayesian Inference under Cluster Sampling with Probability Proportional to Size}
\author[1]{Susanna Makela}
\author[2]{Yajuan Si}
\author[3]{Andrew Gelman}
\affil[1]{\small Department of Statistics, Columbia University}
\affil[2]{\small Survey Research Center, University of Michigan; Corresponding author: yajuan@umich.edu}
\affil[3]{\small Departments of Statistics and Political Science, Columbia University}
\date{2 Oct 2017}

\begin{document}

\maketitle


\begin{abstract}
Cluster sampling is common in survey practice, and the corresponding inference has been predominantly design-based. We develop a Bayesian framework for cluster sampling and account for the design effect in the outcome modeling. We consider a two-stage cluster sampling design where the clusters are first selected with probability proportional to cluster size, and then units are randomly sampled inside selected clusters. Challenges arise when the sizes of nonsampled cluster are unknown. We propose nonparametric and parametric Bayesian approaches for predicting the unknown cluster sizes, with this inference performed simultaneously with the model for survey outcome. Simulation studies show that the integrated Bayesian approach outperforms classical methods with efficiency gains. We use Stan for computing and apply the proposal to the Fragile Families and Child Wellbeing study as an illustration of complex survey inference in health surveys.

\end{abstract}

\section{Introduction}

Cluster sampling has been widely implemented in epidemiology and public health surveys~\citep{review:cluster:Carlin99}. We develop a Bayesian approach for survey inference under cluster sampling, particularly in the absence of design information for nonsampled clusters.  The potential advantages of Bayesian methods are for small area estimation and adjusting for many poststratification factors \citep{gelman07}. However, most of the work in this area has been done for one-stage sampling or ignoring clustering in the data collection; thus it would be a potentially useful contribution to bring cluster sampling into the class of problems for which there are available Bayesian models. In the present paper, we consider two-stage probability proportional to size (PPS) sampling, with the understanding that other designs could be modeled in similar ways.

Cluster sampling increases cost efficiency when partial clusters are included in the probability sampling framework. Bayesian cluster sampling inference is essentially outcome prediction for nonsampled units in the sampled clusters and all units in the nonsampled clusters. The design information should be accounted for in the modeling, but design information for nonsampled clusters is often unknown or inaccessible. We introduce estimation strategies for such information and connect multilevel regression models to sampling design as a unified framework for survey inference.

We consider the design that involves first sampling primary sampling units (PSUs) and then sampling secondary sampling units (SSUs) within selected PSUs. The two-stage cluster sampling design has played an important role in the sequential data collection process of many big health surveys, such as the National Health Interview Survey and the Medical Expenditure Panel Survey. This design requires a complete listing of PSUs and a complete listing of SSUs only within selected PSUs, and thus is widely used when generating a sampling frame of every unit in the population is infeasible or impractical. For example, in designing a nationally representative household survey, generating a complete listing of every household in the country requires essentially as much effort as a complete census of all households. Instead, the sampling proceeds in stages, first sampling PSUs such as counties, cities, or census tracts. The PSUs are sampled with probability proportional to size, which is commonly the number of SSUs in the PSU but can be a more general measure of size, such as annual revenue or agricultural yield. SSUs are then randomly selected within selected PSUs, often with a fixed number or proportion. This design assumes independence and invariance of the second-stage sampling design \citep{sarndal}. Invariance means that the sampling of SSUs is independent of which PSUs are sampled, and independence means sampling of SSUs in one PSU is independent of sampling in other PSUs. For clarification, a two-\textit{phase} design is one in which one or both assumptions do not hold.

Our motivating application survey, the Fragile Families and Child Wellbeing study \citep{ffdesign01}, was collected via a multistage design, where two-stage cluster sampling served as a key step. The study aims to examine the conditions and capabilities of new unwed parents, the wellbeing of their children and the policy and environmental effect. To obtain a nationally representative sample of non-marital births in large U.S. cities, the study sequentially sampled cities, hospitals, and births. The sampling of cities used a stratified random sample of all U.S. cities with 200,000 or more people, where the stratification was based on policy environments and labor market conditions in the different cities. Inside each stratum, cities were selected with probability proportional to the city population size. In the selected cities, all hospitals in the small cities were included, while a random sample of hospitals or the hospital with the largest number of non-marital births was selected in large cities. Lastly, a predetermined number of births were selected inside each hospital. Classical weighting adjustment for the complex study design results in highly variable weights \citep{ffweights08}, thus yielding unstable inferences. We would like to develop a Bayesian framework to account for the complex designs under two-stage cluster sampling.

Our goal is to develop hierarchical models and account for design effects to yield valid and robust survey inference. Bayesian hierarchical models are well-equipped to handle the multistage design and stabilize estimation via smoothing. As an intermediate step, two-stage cluster sampling is crucial in the Fragile Families study to select cities and hospitals. However, cluster sampling presents unique methodology challenges as little information is available on the unselected clusters. This article uses the Fragile Families study as an illustration and focuses on Bayesian cluster sampling inference to build a unified survey inference framework. The unified framework can be extended under a complex sampling design, as discussed in Section~\ref{sec:discussion}.

We illustrate finite population inference with the estimation of the population mean in a two-stage cluster sample. Specifically, we consider a population of $J$ clusters, with each cluster $j$ containing $N_j$ units and a total population size of $N = \sum_{j=1}^J N_j$. Let $I_j$ denote the inclusion indicator for cluster $j$ and $I_{i\mid j}$ denote the inclusion indicator for unit $i$ in cluster $j$, $i = 1, \ldots, N_{j[i]}$, where $j[i]$ denotes the cluster to which unit $i$ belongs. Clusters are sampled with probability proportional to the measure of size $M_j$, which is known to the analyst only for the sampled clusters. Our goal is to estimate the finite population mean of the survey variable $y$, which, for a continuous variable is defined as
\begin{align}\label{eq:ybar_pop}
	\overline{y} = \sum_{j=1}^J \frac{N_j}{N} \overline{y}_j,
\end{align}
where $\overline{y}_j$ represents the mean of $y$ in cluster $j$. For a binary outcome, we seek to estimate the population proportion, which is given by
\begin{align}\label{eq:ybar_prop}
	\overline{y} = \sum_{j=1}^J \frac{y_{(j)}}{N},
\end{align}
where $y_{(j)}$ is the population total in cluster $j$.

Classically, inference in survey sampling has been design-based. The design-based approach treats the survey outcome $y$ as fixed, with randomness arising solely from the random distribution of the inclusion indicator $I$. Design-based estimators have the advantage of being design-consistent, where design-consistency means that the estimator will converge to the true value as the population and sample sizes increase under the given sampling design. However, they are often unstable with large standard errors. For estimating the finite population mean of an outcome $y_i$, the classical design-based estimator for a single-stage sample $s$ of size $n$ is the H\'{a}jek estimator~\citep{sarndal}
$	\widehat{\theta}^{H} = \frac{\sum_{i=1}^n y_i / \pi_i}{\sum_{i=1}^n 1 / \pi_i}$,
where $\pi_i$ is the inclusion (selection and response) probability of unit $i$. In the two-stage sample $s$, when $J_s$ out of $J$ clusters are selected with $n_j$'s sampled SSUs, for convenience labeled as $j=1,\dots,J_s$, the estimator becomes
\begin{align}\label{eq:hajek_2s}
	\widehat{\theta}^{H} = \frac{\sum_{j=1}^{J_s} \left( \sum_{i=1}^{n_j} y_i / \pi_{i \mid j} \right) / \pi_j}{\sum_{j=1}^{J_s} N_j / \pi_j},
\end{align}
where $\pi_j$ is the selection probability of cluster $j$, and $\pi_{i \mid j}$ is the selection probability of unit $i$ in cluster $j$ given that cluster $j$ was sampled.

The design-based approach does not require a statistical model for the survey outcomes but implicitly assumes the specific outcome model structure and linearity~\citep{samsi:review17}. The performance relies on the validity of the model assumptions, and then the estimators can yield biased inference under invalid assumptions. Another major challenge with design-based estimators comes in estimating their variance. The variance of a design-based estimator generally requires knowledge of not only the inclusion probability $\pi_i$ for a given unit $i$, but also the joint inclusion probability $\pi_{ii'}$ for any two units $i$ and $i'$ in the population. This information is often unavailable in practice, such as the unknown measure of size for nonsampled clusters under the PPS setting. Joint inclusion probabilities can be challenging to compute even for straightforward sampling designs, and variance estimators for design-based estimators are often based on simplifications and approximations. Furthermore, the inverse-probability of inclusion weighting often leads to highly variable estimators.

Bayesian inference, in contrast, directly models both the inclusion indicators $I_i$ and the survey outcomes $y_i$. The Bayesian approach to survey inference has many advantages over the design-based approach, including the ability to handle complex design features such as multistage clustering and stratification, stabilized inference for small-sample problems, incorporation of prior information, and large-sample efficiency~\citep{little2004}. When the design variables are included in the model, the selection mechanism becomes ignorable~\citep{rubin83-pi,bda3}, and we can model the outcomes $y$ alone, instead of jointly modeling $y$ and the inclusion indicator $I$. The importance of including design variables in the model has also been emphasized for missing data imputation~\citep{Schafer97,reiter:cluster06}.

Unfortunately, in many (arguably most) practical situations, the set of design variables is not available for the entire population and is instead known only for sampled clusters or units. In the case of PPS sampling, where the design variables consist of the cluster measures of size $\{M_j\}_{j=1}^J$, we as the survey analysts may only have access to $M_j$ (or, equivalently, the inclusion probability $\pi_j$) for the sampled subset of $J_s$ clusters. This missing information on measures of size causes methodology challenge in the Bayesian setting because we cannot predict the values of $y$ for the nonsampled clusters without it. We need to model the values of $M_j$ for nonsampled clusters before we are able to make inferences about $\overline{y}$ conditional on the design information.

Existing Bayesian approaches to this problem \citep{zangeneh2011, zangeneh2015} consider the case of single-stage PPS sampling. In addition, they separate estimation of the missing measure sizes and inference for the finite population quantities into two steps. In contrast, we propose an approach that integrates these steps into one model for a two-stage cluster sample. Our model allows for both cluster- and unit-level information to be used when both are available in certain cases. For much of this paper, we assume the measure of size is equal to the cluster size $N_j$ and use $N_j$ in place of $M_j$ for simple illustration. 

The rest of this paper proceeds as follows. Section~\ref{sec:methods} first gives an overview of current approaches to finite population inference under PPS and then describes our approach and its advantages. In Section~\ref{sec:simulation}, we describe a simulation study to investigate the performance of our method and compare with other literature methods. We apply our proposal to the Fragile Families study in Section~\ref{sec:application} and discuss the conclusions and extensions in Section~\ref{sec:discussion}.

\section{Methods}
\label{sec:methods}

In the two-stage cluster sampling, a fixed number $J_s$ of clusters are sampled with PPS, so that the probability of cluster $j$ being included in the sample is proportional to $N_j$: 
\[\textrm{Pr}(I_j = 1 \mid N_j) \propto N_j.\]
 We only observe $N_j$'s for the clusters in the sample, that is, the empirical distribution of $(N_j|I_j=1)$. Our proposed procedure simultaneously models the population cluster sizes and the outcome and propagates the estimation uncertainty. 

Let $x_i$ denote the auxiliary variables that are predictive for the outcome. The observed data are $(y_{obs}, x_{obs}, N_{obs}, \overline{x}_{1:J}, N, J, J_s)$, where $\overline{x}_{1:J}$ is the cluster-level mean of the covariate $x$ for all clusters $j = 1, \ldots, J$, and $N$, $J$, and $J_s$ are the total population size, total number of clusters, and number of sampled clusters, respectively. The subscript $obs$ denotes the observed portions of the variables: $y_{obs} = \{ y_i: i = 1, \ldots, n_{j[i]}, j = 1, \ldots, J_s \}$, $x_{obs} = \{ x_i: i = 1, \ldots, n_{j[i]}, j = 1, \ldots, J_s \}$, $N_{obs} = \{ N_j: j = 1, \ldots, J_s \}$, where for convenience we number the sampled clusters $j = 1, \ldots, J_s$ and the nonsampled clusters as $j = J_s + 1, \ldots, J$. We assume that $x_i$ is known for all sampled units, and that $\overline{x}_j$ is known for all clusters. If $x$ is a demographic covariate, in practice it's often the case that we know demographic characteristics of clusters even if the cluster size is unknown.

The goal is to estimate the finite population mean $\overline{y}$, defined for a continuous outcome,
\begin{align*}
	\overline{y} &= \sum_{j=1}^J \frac{N_j}{N} \overline{y}_j
	= \frac{1}{N} \left( \sum_{j=1}^{J_s} \frac{n_j \overline{y}_{obs,j} + (N_j - n_j) \overline{y}_{exc,j}}{N_j} + \sum_{j=J_s+1}^J N_{exc,j} \overline{y}_{exc, j}  \right),
\end{align*}
where $\overline{y}_{obs,j}$ is the mean of the sampled units in sampled cluster $j$, $\overline{y}_{exc,j}$ is the mean of the nonsampled units in  cluster $j$, and $N_{exc,j}$ is the size of nonsampled cluster $j$ that is unknown. For a binary outcome, the population proportion is
\begin{align*}
	\overline{y} &= \sum_{j=1}^J \frac{y_{(j)}}{N}
	= \frac{1}{N} \left( \sum_{j=1}^{J_s} \left( y_{obs,(j)} + y_{exc,(j)} \right) + \sum_{j=J_s+1}^J y_{exc,(j)} \right),
\end{align*}
where $y_{(j)}$ is the total of all units in cluster $j$, $y_{obs,(j)}$ is the total of sampled units in sampled cluster $j$ and $y_{exc,(j)}$ is the total of the binary outcome in nonsampled units in cluster $j$.

We assume the continuous survey outcome $y$ is related to the covariate $x$ and cluster sizes $N_j$ in the following way:
\begin{align}
	\label{eq:main_model_y} y_i &\sim \textrm{N}(\beta_{0j[i]} + \beta_{1j[i]} x_i, \sigma_y^2)  \\
	\label{eq:main_model_beta0} \beta_{0j} &\sim \textrm{N}(\alpha_0 + \gamma_0 \log^c(N_j), \sigma_{\beta_0}^2)\\
	\label{eq:main_model_beta1} \beta_{1j} &\sim \textrm{N}(\alpha_1 + \gamma_1 \log^c(N_j), \sigma_{\beta_1}^2) \\
	\label{eq:main_model_Nj} N_j &\sim p(N_j \mid \phi), 
\end{align}
where $\phi$ are the parameters governing the distribution of the cluster sizes $N_j$. The model assumes the regression coefficients are cluster-varying and depend on the cluster sizes. We use random-effects model to borrow information across clusters. While fixed-effects model with cluster membership indicators can also be used to quantify the cluster effect, fixed cluster effects models may increase the variance, as shown by~\cite{reiter:cluster06} and \cite{andridge:cluster11} in the context of missing data imputation. In addition, predictions cannot be made for nonsampled clusters using fixed-effects models since no units are available. 

Our model for a binary outcome is identical, except that we modify \eqref{eq:main_model_y} to be
\begin{align}\label{eq:bin_model_y}
	\textrm{Pr}(y_i = 1) = \textrm{logit}^{-1}(\beta_{0j[i]}),
\end{align}
and omit \eqref{eq:main_model_beta1}. We exclude unit-level covariates in the binary case because the nonlinear nature of the inverse logit link function makes it challenging to make use of data at the unit level. Specifically, predicting $\overline{y}_{exc,j}$ requires knowing $x_i$ for all nonsampled units in cluster $j$, and if we knew this, clearly we would have also known $N_j$ for nonsampled clusters $j$. 

We use the centered logarithms of the cluster sizes $\log^c(N_j)$ as predictors; we work on the logarithm scale to better accommodate large cluster sizes and center for interpretation convenience. The sampling is assumed to be ignorable after including the design variables in the outcome model. 

We assign an estimation model $p(N_j \mid \phi)$ to the cluster sizes, which we observe only for the sampled clusters. We develop both nonparametric and parametric modeling strategies to predict the cluster sizes of nonsampled clusters.

We use $\psi$ to denote the regression parameters $\psi = (\alpha_0, \gamma_0, \sigma_0, \alpha_1, \gamma_1, \sigma_1, \sigma_y)$ associated with the outcome modeling and $\theta$ for all parameters of interest: $\theta = (\psi, \phi)$. The likelihood for the observed data is
\[
	p(y_{obs} \mid x_{obs}, N_{obs}, \theta) \propto p(y_{obs} \mid x_{obs}, N_{obs}, \psi) p(N_{obs} \mid \phi),
\]
and the posterior distribution is
\[
	p(\theta \mid y_{obs}, x_{obs}, N_{obs}) \propto p(y_{obs} \mid x_{obs}, N_{obs}, \psi) p(N_{obs} \mid \phi) p(\psi)p(\phi),
\]
where we assume that $\psi$ and $\phi$ are independent, allowing us to write $p(\theta) = p(\psi)p(\phi)$. 

Because of the independence, invariance and ignorability assumptions in the two-stage cluster sampling, the distribution of the outcome $y$, given the design variables, is the same in the sample and the population; that is, the observed data likelihood is the same as the complete data likelihood,
\[
	p(y_{obs} \mid x_{obs}, N_{obs}, \psi) = p(y \mid x, N, I=1, \psi) = p(y \mid x, N, \psi).
\]
Here $p(y \mid x, N, \psi)$ is specified by \eqref{eq:main_model_y}--\eqref{eq:main_model_beta1} for continuous $y$ and by \eqref{eq:bin_model_y} and \eqref{eq:main_model_beta0} for binary $y$.

The challenge lies in estimating the distribution of the $N_j$'s when the sampling is informative. Under PPS sampling, the probability of observing a cluster of size $N_j$ is
\begin{align}\label{eq:obsNj_dist}
	p(N_j \mid I_j = 1) &\propto \textrm{Pr}(I_j = 1 \mid N_j) p(N_j) \nonumber \\
	&\propto N_j p(N_j).
\end{align}
We consider both nonparametric and parametric modeling strategies for the prior distribution $p(N_j)$ (also called the population distribution, to distinguish from the distributions of sampled and nonsampled cluster sizes) in~\eqref{eq:main_model_Nj}. First, we introduce the Bayesian bootstrap algorithm in Section~\ref{sec:bb} as a nonparametric approach to predicting the unobserved $N_j$'s. Second, we investigate two parametric distributional assumptions in Section~\ref{sec:sb_dists} for $p(N_j)$, the negative binomial and lognormal distributions. Here our goal is to directly model the distribution of the cluster sizes accounting for the fact that the observed distribution is biased from the complete population distribution. Following \cite{pr1978}, we refer to these parametric choices as size-biased distributions.

\subsection{Bayesian bootstrap}\label{sec:bb}

For a nonparametric model of the sampled cluster sizes, we implement the Bayesian bootstrap algorithm in \cite{littlezheng2007} and \cite{zangeneh2011} for one-stage PPS sampling and modify it under two-stage PPS sampling. Without a parametric assumption for $p(N_j)$, we connect $p(N_j \mid I_j = 0)$ with $p(N_j \mid I_j = 1)$ through the empirical distributions under PPS sampling. Assume the $N_j$'s observed for the sampled clusters have $B$ unique values $N_1^*,\dots, N_B^*$, and let $k_1,\dots, k_B$ be the corresponding counts of these unique sizes, such that $\sum_b k_b = J_s$. Let $\psi_b$ denote the probability of observing a cluster of size $N_b^*$ in the sample: $\psi_b = \textrm{Pr}(N_j=N_b^*\mid I_j=1)$. We can then model the counts $k=(k_1,\dots,k_B)$ as multinomially distributed with total $J_s$ and parameters $\psi=(\psi_1,\dots,\psi_B)$. The observed likelihood $L_{obs}(\psi)$ is,
\[
\textrm{Pr}\!\left(\!\left. k_1 =\! \sum_{j=1}^{J_s} I(N_j=N_1^*), \dots, k_B =\! \sum_{j=1}^{J_s} I(N_j=N_B^*) \right| I_j=1, j=1,\dots, J_s\right) \propto \prod_{b=1}^B\psi_b^{k_b},
\]
where $I(\cdot)$ is an indicator function, $I(\cdot)=1$ if the inside expression is true and $0$ otherwise. The $\psi$'s are given a noninformative Haldane prior: $p(\psi_1, \ldots, \psi_B) = \textrm{Dirichlet}(0, \ldots, 0)$, a conjugate Dirichlet prior distribution. The posterior distribution of $\psi$ is then
\[
	p(\psi_1, \ldots, \psi_B | k_1, \ldots, k_B)	= \textrm{Dirichlet}(k_1, \ldots, k_B).
\]

Suppose the unique values of $N_j$'s cover all possible values in the population. Assume $k^{\star}_b$ is the number of nonsampled clusters with size $N^*_b$, for $b=1,\dots,B$, and let $\psi^{\star}_b$ denote the probability of an unobserved cluster having size $N_b^*$: $\psi^{\star}_b=\textrm{Pr}(N_j=N_b^*\mid I_j=0)$. Then the counts of the $B$ unique sizes among the nonsampled clusters, $(k^{\star}_1,\dots,k^{\star}_B)$, follow a multinomial distribution with total $J-J_s=\sum_bk^{\star}_b $ and probabilities $(\psi^{\star}_1,\dots,\psi^{\star}_B)$:
\[
	p(k^{\star}_1,\dots,k^{\star}_B \mid J - J_s, \psi^{\star}_1,\dots,\psi^{\star}_B) \propto \prod_{b=1}^B \psi_b^{\star k^{\star}_b}
\]
Using Bayes' rule, we can write $\psi^{\star}_b$ as
\begin{align*}
\psi^{\star}_b&=\textrm{Pr}(N_j=N_b^*\mid I_j=0)\nonumber\\
	&\propto \textrm{Pr}(N_j=N_b^*\mid I_j=1)\frac{\textrm{Pr}(I_j=0|N_j=N_b^*)}{\textrm{Pr}(I_j=1|N_j=N_b^*)}\nonumber\\
	&=\psi_b\frac{1-\pi_b}{\pi_b},
\end{align*}
where $\pi_b=\textrm{Pr}(I_j=1|N_j=N_b^*)=J_s N_b^*/N$ is the conditional cluster selection probability known in the PPS sample, $J_s$ is the number of sampled clusters, and $N$ is the population size. This approach essentially adjusts the probability of resampling an observed size $N^*_b$ by the odds of a cluster of that size not being sampled, so that smaller sizes are upweighted relative to larger ones. 

Given the posterior draws of $\psi^{\star}_b$'s and $k^{\star}_b$'s, we create $k^*_b$ replicates of the size $N^*_b$, yielding a sample of the nonsampled cluster sizes from their posterior predictive distribution. The Bayesian bootstrap for cluster sampling is similar to the ``two-stage P\'olya posterior" approach proposed by \cite{Meeden99}, which simulates draws that form a population of clusters and then an entire population of elements within each cluster. \cite{mi2stage:zhou16} incorporated weights in Bayesian bootstrap for multiple imputation in two-stage cluster samples. \cite{si2015} uses a similar approach to estimating the poststrafication cell sizes constructed by the survey weights.

The Bayesian bootstrap avoids parametric assumption on the population distribution $p(N_j)$ and use the empirical distribution in the observed clusters. This implicitly introduces a noninformative prior distribution on $N_j$'s. However, this approach restricts the draws for the nonsampled cluster sizes to come from the set of observed cluster sizes, where small clusters may be omitted under PPS sampling. While the Bayesian bootstrap is a robust algorithm for predicting the unknown $N_j$'s, we can achieve efficiency gains with a parametric distribution on $p(N_j)$, especially when prior distribution information is available.

\subsection{Size-biased distributions}\label{sec:sb_dists}

Inducing parametric sized-biased distributions follows the superpopulation concept in the model-based survey inference literature and incorporates informative prior information. In practice, we may have some knowledge about the cluster sizes, such as the distribution in a similar population or from previous years. We can incorporate this additional information through the prior distribution specification. Sized-biased distributions were discussed by \cite{pr1978} for population size estimation. We consider a discrete and a continuous distribution as candidates for modeling the size distributions. The observed likelihood is connected with the proposed population distribution via \eqref{eq:obsNj_dist}. Using the PPS sample, we can estimate the parameters in the population distribution and then predict the nonsampled cluster sizes.

For the discrete case, we assume the population cluster sizes $N_j$ follow a negative binomial distribution: $N_j \sim \textrm{NegBin}(k,p)$, with $k > 0$ and $p \in (0, 1)$. By normalizing the distribution in \eqref{eq:obsNj_dist} and completing the algebra shown as below, we see that the sizes in the PPS sample can be written as $N_j = 1 + W_j$, where $W_j \sim \textrm{NegBin}(k+1, p)$. 

For $m = 0, 1, 2, \ldots$, the probability of observing $N_j=m$ in the PPS sample is
\begin{align*}
	\textrm{Pr}(N_j = m \mid I_j = 1) 
	&=\frac{\textrm{Pr}(I_j = 1 \mid N_j = m) \textrm{Pr}(N_j = m)}{\textrm{Pr}(I_j = 1)}\\[5pt]
	&= \frac{m \binom{m + k - 1}{m} p^k (1-p)^m}{\sum_{m=0}^\infty m \binom{m + k - 1}{m} p^k (1-p)^m} \\[5pt]
	&= \binom{(m-1)+(k+1)-1}{m-1} p^{k+1} (1-p)^{m-1}	\\[5pt]
	&= \textrm{Pr}(W=m-1),
\end{align*}
where $W \sim \textrm{NegBin}(k+1, p)$. 

For the continuous case, we use the lognormal distribution. If the population distribution is $N_j \sim \textrm{lognormal}(\mu, \tau^2)$, then $(N_j\mid I_j=1) \sim \textrm{lognormal}(\mu + \tau^2, \tau^2)$. To see this, recall that $p(N_j)$ denotes the pdf of size variables $N_j$ in the population. Then the pdf of $N_j$ in the PPS sample is
\begin{align}\label{eq:ln_pt1}
	p(N_j \mid I_j = 1) &=\frac{\textrm{Pr}(I_j = 1 \mid N_j) p(N_j )}{\textrm{Pr}(I_j = 1)}  \nonumber\\[5pt]
	&= \frac{(\sqrt{2\pi}\tau)^{-1} \exp \left( -\frac{(\log N_j - \mu)^2}{2 \tau^2} \right) }{\int_0^{\infty} (\sqrt{2\pi}\tau) ^{-1}\exp \left( -\frac{(\log N_j - \mu)^2}{2 \tau^2} \right) dN_j} \nonumber \\[5pt]
	&=  \frac{\exp \left( -\frac{(\log N_j - \mu)^2}{2 \tau^2} \right) }{\int_0^{\infty} \exp \left( -\frac{(\log N_j - \mu)^2}{2 \tau^2} \right) dN_j}.
\end{align}
We can now simplify the denominator:
\begin{align}\label{eq:ln_pt2}
	\int_0^{\infty} &\exp \left( -\frac{(\log N_j - \mu)^2}{2 \tau^2} \right) dN_j 
	= \sqrt{2\pi}\tau \exp \left( \mu + \frac{\tau^2}{2} \right).
\end{align}
Now, substitute \eqref{eq:ln_pt2} for the denominator in \eqref{eq:ln_pt1}:
\begin{align*}
	p(N_j \mid I_j = 1) &= \frac{1}{\sqrt{2\pi}\tau} \exp \left( -\frac{(\log N_j - \mu)^2}{2 \tau^2} - (\mu + \frac{\tau^2}{2}) \right) \\[5pt]
	&= \frac{1}{N_j\sqrt{2\pi}\tau} \exp \left( -\frac{(\log N_j - (\mu + \tau^2))^2}{2 \tau^2} \right).
\end{align*}
Thus, the distribution of sampled cluster sizes in the PPS sample is $(N_j|I_j=1) \sim \textrm{lognormal}(\mu +\tau^2, \tau^2)$. 

Regardless of the parametric model we choose, in order to generate predictions of the nonsampled cluster sizes, we need to draw from $p(N_j \mid I_j = 0)$. We apply rejection sampling and use samples from $p(N_j)$ to approximate the sampling from $p(N_j \mid I_j = 0)$.
\[
p(N_j \mid I_j = 0)=\frac{\textrm{Pr}(I_j=0\mid N_j)p(N_j)}{\textrm{Pr}(I_j=0)}\triangleq Gp(N_j),
\]
where $G\triangleq\textrm{Pr}(I_j=0\mid N_j)/\textrm{Pr}(I_j=0)$ has a constant upper bound shown as below. The marginal probability selection for cluster $j$ is $\textrm{Pr}(I_j = 1) = J_s/J$, and the joint distribution of $(N_j, I_j)$ in the PPS sample is $p(N_j, I_j=1)=cN_jp(N_j)$, where $c$ is a constant. And

\[\textrm{Pr}(I_j = 1)=\int_{N_j} p(N_j, I_j=1)dp(N_j)=\int_{N_j} cN_jp(N_j)dp(N_j)=c~\mbox{E}(N_j).\] Hence, $c = J_s/(J\mbox{E}(N_j))$. Then
\begin{align*}
G = \frac{1-\textrm{Pr}(I_j=1\mid N_j)}{1-\textrm{Pr}(I_j=1)}=\frac{1-\frac{J_sN_j}{J\mbox{E}(N_j)}}{1-J_s/J}.
\end{align*}
\cite{zangeneh2011} assumes that $\mbox{E}(N_j)=N / J$, approximated by the finite sample cluster size, such that
\begin{align*}
G= \frac{1-\frac{J_sN_j}{N}}{1-J_s/J}\leq \frac{J}{J-J_s}.
\end{align*}
Given the posterior distribution of $p(N_j \mid -)$, we use rejection sampling to obtain posterior predictive samples from $p(N_j\mid I_j=0, -)$.

\subsection{Prior specification and computation}\label{sec:prior}

We use the following weakly informative prior distributions as recommended by \cite{gelman06-prior},
\begin{align*}
	\alpha_0, \gamma_0, \alpha_1, \gamma_1 &\overset{ind}{\sim} \textrm{N}(0, 10) \\
	\sigma_{\beta_0}, \sigma_{\beta_1}, \sigma_y &\overset{ind}{\sim}\textrm{Cauchy}^+(0, 2.5). 
\end{align*}
Here $\textrm{Cauchy}^+(0, 2.5)$ denotes a Cauchy distribution with location 0 and scale 2.5 restricted to positive values. The weakly informative prior specification will allow the group-level variance parameters to be close to 0 and have large tail values.

For the parameters governing the distribution of $N_j$, such as $(k,p)$ in the negative binomial distribution or $(\mu,\tau)$ in the lognormal distribution, we can use noninformative priors when the number of clusters sampled is large. However, when only a few clusters are sampled, we need informative prior information to counteract the sparsity of the data and stabilize the inference. This is particularly true when using a model for the cluster sizes that includes implicit assumptions about the data. For example, as an overdispersed extension of the Poisson distribution, the negative binomial distribution assumes that the data come from a distribution whose mean is smaller than the variance. However, in a sample of only five clusters, it may well be that the sample mean is larger than the sample variance, making it difficult to fit the negative binomial distribution to the data without strong prior information. In this case, we reparameterize the negative binomial as a Gamma mixture of Poisson distributions and place a prior on the coefficient of variation (CV), the standard deviation divided by the mean. In this case, the CV works out to the reciprocal of the square root of the scale parameter of the Gamma distribution. With a small number of clusters, we expect the CV to be close to one and therefore use an exponential prior distribution with rate 1. For the lognormal distribution, we place a $\textrm{Cauchy}^+(0, 2.5)$ prior on the scale parameter $\tau$. To aid estimation for the case with only a few sampled clusters, we standardize the log of the sampled cluster sizes by subtracting their mean and dividing by the standard deviation. 

For the continuous outcome, in nonsampled clusters $j$, the posterior predictive distribution for $\overline{y}_{exc,j}$ is
\[
	(\overline{y}_{exc,j} \mid \cdot) \sim \textrm{N} \left( \beta_{0j} + \beta_{1j} \overline{x}_j, \sigma_y^2 / N_j \right),
\]
where we assume $\overline{x}_j$ is known. Specifically, we draw new values of $\beta_{0j}$, $\beta_{1j}$, $\sigma_y$, and $N_j$ from their posterior distributions and then draw $\overline{y}_{exc,j}$ from the above distribution. In sampled clusters, the posterior predictive distribution for the nonsampled units is 
\[
(\overline{y}_{exc,j} \mid \cdot)  \sim N \left( \beta_{0j} + \beta_{1j} \overline{x}_j, \sigma_y^2 / (N_j-n_j) \right).
\] 
When $N_j$ is large compared to $n_j$, as is the case in many large-scale surveys and specifically in the Fragile Families study, $\overline{y}_{exc,j}$ is close to the cluster mean $\overline{y}_j$ and is well approximated by $\beta_{0j} + \beta_{1j} \overline{x}_j$, which we calculate using the posterior means of $\beta_{0j}$ and $\beta_{1j}$.

The posterior computation is implemented in Stan \citep{stan}, which conducts full Bayesian inference and generates the posterior samples. The estimation for the outcome model and the cluster size model can be integrated into the posterior computation, which allows for uncertainty propagation throughout the parameter estimates, in contrast to previous approaches \cite[e.g.,][]{littlezheng2007,zangeneh2015}. 

Stan is unique in providing detailed warnings and diagnostics to inform the user when posterior inferences may be unreliable due to difficulties in sampling. Divergent transitions indicate that the sampler is unable to explore a portion of the parameter space, which can lead to significant bias in the resulting posterior distribution and ultimately unreliable inferences \citep{stan_manual}. Stan reports the number of divergent transitions for each run, and even one divergent transition indicates that the results may be suspect. If divergent transitions occur, we follow the recommendation of Stan developers and iteratively increase the target acceptance rate \texttt{adapt\_delta} \citep{stan_warnings}. If divergent transitions occur even with $\texttt{adapt\_delta} = 0.99999$, we switch to the noncentral parameterization and follow the same procedure for increasing adapt delta as necessary. The noncentral parameterization is a mathematically equivalent formulation for the model that can avoid posterior geometries that are difficult for HMC to explore; see \cite{betancourt_girolami} and \cite{stan_manual}. 

To understand the importance of explicitly controlling for all design variables in this context, we also fit a model similar to \eqref{eq:main_model_y}--\eqref{eq:main_model_Nj} but with $\gamma_0$ and $\gamma_1$ set to 0. Such a model accounts for the hierarchical cluster nature of the data by allowing $\beta_0$ and $\beta_1$ to vary by cluster, but does not account for the PPS sampling design since the cluster sizes $N_j$ are excluded from the model:
\begin{align}\label{eq:cluster_inds_only}
	y_i &\sim \textrm{N}(\beta_{0j[i]} + \beta_{1j[i]} x_i, \sigma_y^2) \quad \quad \textrm{(continuous)} \nonumber \\
	\textrm{Pr}(y_i=1) &= \textrm{logit}^{-1}(\beta_{0j[i]}) \quad \quad \textrm{(binary)} \nonumber \\
	\beta_{0j} &\sim \textrm{N}(\alpha_0, \sigma_{\beta_0}^2) \\
	\beta_{1j} &\sim \textrm{N}(\alpha_1, \sigma_{\beta_1}^2) \nonumber 
\end{align}

\section{Simulation study}
\label{sec:simulation}

We perform a simulation study to compare the performance of our integrated approaches with classical design-based estimators on the statistical validity of the finite population inference. We generate a population from which we take repeated two-stage cluster samples under PPS and use each of the methods to estimate $\overline{y}$. The population consists of $J=100$ clusters, with cluster sizes $N_j$ drawn from one of two distributions. The first is a Poisson distribution with rate 500. The second is a multinomial distribution over scaled Gamma-distributed sizes. Specifically, we draw $J=100$ candidate cluster sizes $N_j$ as $N_j = 100G_j$, where $G_j \sim \text{Gamma}(10, 1)$. We then take a multinomial draw from these 100 unique sizes, with the $J$-vector of probabilities drawn from a Dirichlet distribution with concentration parameter 10, which disperses probability mass equally across the $J=100$ components. In both cases, to avoid clusters that would be selected with probability 1, we resample the $J$ cluster sizes until none are so large to be selected with certainty.

For continuous outcome, we simulate the population unit value $y_i$ from the following:
\begin{align}\label{eq:dgp}
	y_i &\sim \textrm{N}(\beta_{0j[i]} + \beta_{1j[i]} x_i, \sigma_y^2) \nonumber \\
	\beta_{0j} &\sim \textrm{N}(\alpha_0 + \gamma_0 \log^c(N_j), \sigma_{\beta_0}^2) \nonumber \\
	\beta_{1j} &\sim \textrm{N}(\alpha_1 + \gamma_1 \log^c(N_j), \sigma_{\beta_1}^2) \\
	\alpha_0, \alpha_1, \gamma_0, \gamma_1 &\sim \textrm{N}(0, 1) \nonumber \\
	\sigma_{\beta_0}, \sigma_{\beta_1}  &\sim \textrm{N}^+(0, 0.5) \nonumber \\
	\sigma_y &\sim \textrm{N}^+(0, 0.75) \nonumber,
\end{align}
where $\textrm{N}^+(\mu, \sigma)$ denotes the positive part of the normal distribution with mean $\mu$ and standard deviation $\sigma$. The model for binary $y$ is identical, except that the first line of \eqref{eq:dgp} is replaced with $y_i \sim \textrm{Bern}(\textrm{logit}^{-1} (\beta_{0j[i]}))$ (and we omit $\beta_{1j}$). 

We use the same outcome model for data generation and estimation to focus on the performance evaluation of different approaches accounting for the design effect and avoid potential model misspecification. In practice the outcome model can be adapted with flexible choices, as discussed in Section~\ref{sec:discussion}. We generate $x_i$ by sampling from the discrete uniform distribution between 20 and 45 and center it by subtracting the mean. We assume that $x_i$ is known for all sampled units, and that $\overline{x}_j$ is known for all clusters. The cluster sizes $N_j$'s are only known in the sampled clusters.

We sample $J_s<J$ clusters using random systematic PPS sampling with probability proportional to the cluster size $N_j$ and $n_j$ units via SRS in each selected cluster $j$. We consider values of $J_s \in \{10, 50\}$ and $n_j \in \{0.1N_j, 0.5N_j, 10, 50\}$. Note that when $n_j \in \{10, 50\}$, the sample is self-weighting, meaning each unit has an equal probability of selection. To see this, recall that the probability of sampling cluster $j$ is $\pi_j \propto N_j$. Since within-cluster sampling is done with SRS, the probability of sampling unit $i$ given cluster $j$ is selected is $\pi_{i \mid j} = n_j / N_j = n / N_j$ when $n_j$ is the same for all clusters. The marginal probability of sampling unit $i$ is therefore $\pi_i = \pi_j \pi_{i \mid j} \propto N_j \cdot (n / N_j) = n$, which is constant across units and clusters. Even though the final weights are constant, our studies show that the design features should be accounted in the outcome model.

For each combination of $J_s$ and $n_j$, we draw $100$ two-stage samples from the population. For each sample, we estimate the finite population mean using the methods described below. 
\begin{itemize}
	\item \texttt{negbin}: The negative binomial size-biased distribution as described in Section \ref{sec:methods};
	\item \texttt{lognormal}: The lognormal size-biased distribution as described in Section \ref{sec:methods};
	\item \texttt{bb}: The Bayesian bootstrap as described in Section \ref{sec:methods};
	\item \texttt{H\'ajek}: The H\'{a}jek estimator in \eqref{eq:hajek_2s};
	\item \texttt{greg}: The generalized regression estimator \citep{greg92}, which leverages a unit-level covariate to improve prediction. We only use this estimator for continuous $y$. To estimate the variances of the H\'{a}jek and generalized regression estimators, we use the formulas given in Chapter 8 of \cite{sarndal};\footnote{In some cases, the sample size is so large as to make calculating the design-based variance under a non-self-weighting design difficult. This is due to the $\check{\Delta}_{k\ell}$ term in equations 8.6.3 and 8.9.27 in \cite{sarndal}, which requires generating an $n \times n$ matrix, where $n = \sum_{j=1}^{J_s} n_j$. When $J_s = 50$ and $n_j = 0.5N_j$, $n$ can easily be 20000 or larger, making the matrix prohibitively large to compute. In these cases, we estimate the variance by randomly selecting 100 units via SRS in each sampled cluster and using those units to compute the required matrix.}
	\item \texttt{cluster\_inds}: The model in \eqref{eq:cluster_inds_only}, which accounts for the hierarchical nature of the data via random cluster effects but does not use the cluster sizes as a cluster-level predictor in modeling $\beta_{0j}$ and $\beta_{1j}$;
	\item \texttt{knowsizes}: The model in \eqref{eq:main_model_y}--\eqref{eq:main_model_beta1}, where we additionally assume the cluster sizes are known for the entire population. This is the best scenario and will serve as a benchmark for the other Bayesian methods.
\end{itemize}

There are three main comparisons that we make in evaluating the results of the simulation study. First, we measure the performance of our proposed integrated Bayesian approach against that of the classical design-based estimators; we do this by comparing the performance of \texttt{negbin}, \texttt{lognormal}, and \texttt{bb} to that of \texttt{H\'ajek} and \texttt{greg}. Second, among the Bayesian methods, we want to understand when the parametric models \texttt{negbin} and \texttt{lognormal} outperform the nonparametric Bayesian bootstrap \texttt{bb}. Third, we compare the performances of \texttt{cluster\_inds} and \texttt{knowsizes} in order to understand the importance of explicitly including cluster sizes as cluster-level predictors in \eqref{eq:main_model_beta0} and \eqref{eq:main_model_beta1}. In this case, we assume that cluster sizes are known for all clusters in the population and focus on the effects of incorrectly excluding or including the cluster sizes as cluster-level predictors in the model.

We carefully monitor the diagnostics of computation performance for each drawn sample. If divergent transitions remain, we discard the sample. We monitor the estimated potential scale reduction factor $\widehat{R}$ for each parameter. This diagnostic assesses the mixing of the chains; at convergence, $\widehat{R} = 1$. If $\widehat{R} \geq 1.1$ for any parameter, we increase the number of iterations by 1000 until all values of $\widehat{R}$ are less than 1.1, up to 4000 iterations. If values of $\widehat{R} \geq 1.1$ remain with 4000 iterations, we discard the drawn sample. The results presented here are based on a minimum of 85 simulation draws for each combination of number of clusters sampled and number of units sampled. That is, we repeatedly draw 100 samples from the population and keep the $L$ cases with good computation performance, $85 \leq L \leq 100$.

\begin{figure}
	\centering
	\centerline{\includegraphics[scale=0.5]{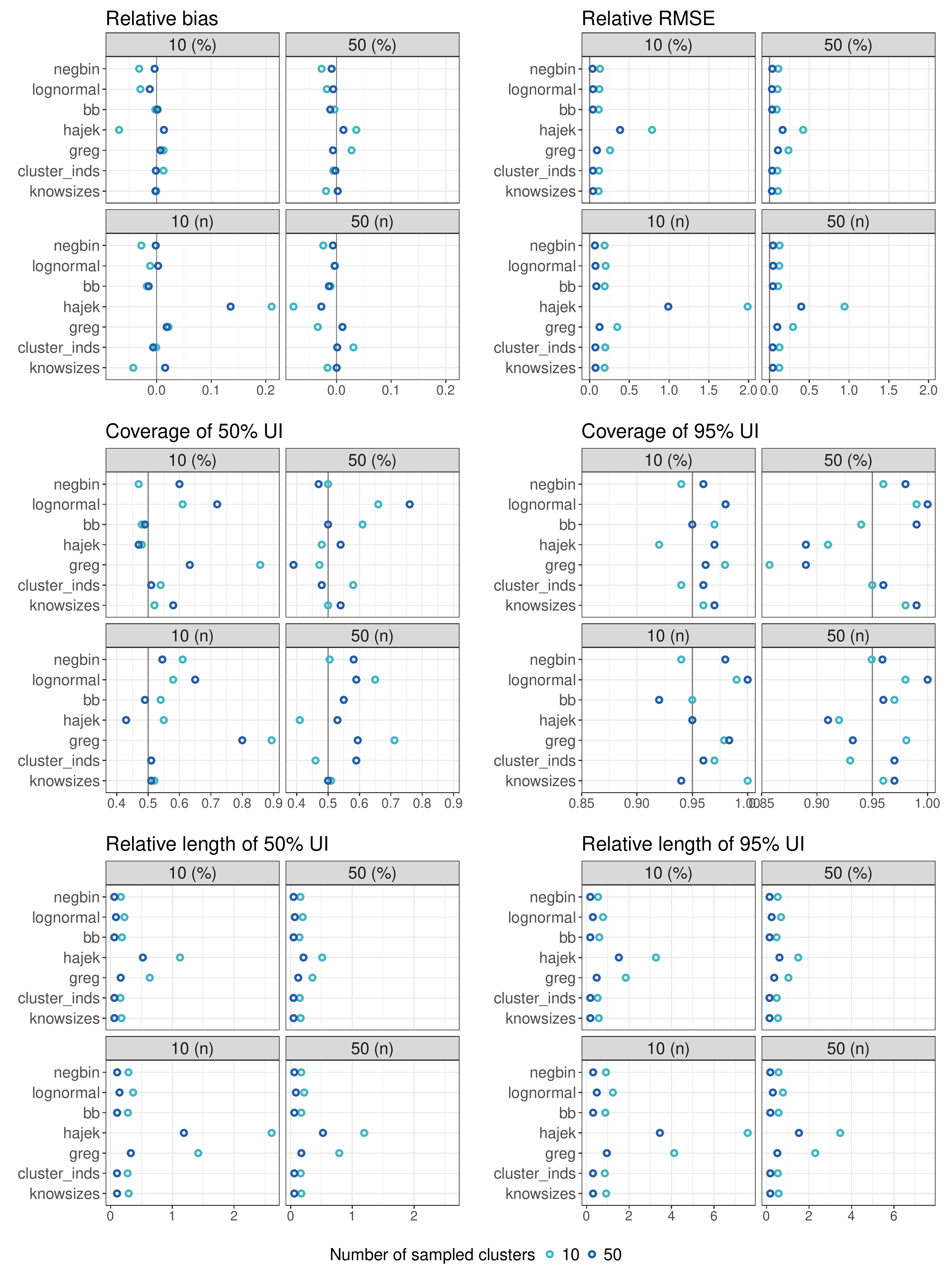}}
	\caption{\em \small Results for continuous $y$ with cluster sizes $N_j$ drawn from a Poisson distribution. The top two plots are for the fixed-percentage SRS schemes, and the bottom two are for the fixed-number SRS samples. \texttt{negbin}: negative binomial distribution; \texttt{lognormal}: lognormal distribution; \texttt{bb}: Bayesian bootstrap; \texttt{H\'ajek}: the H\'{a}jek estimator; \texttt{greg}: generalized regression estimator; \texttt{cluster\_inds}: the model with random cluster effects but without the cluster size predictor; \texttt{knowsizes}: the model with known population cluster sizes.}
	\label{fig:ybar_cont_pois}
\end{figure}

\begin{figure}
	\centering
	\centerline{\includegraphics[scale=0.5]{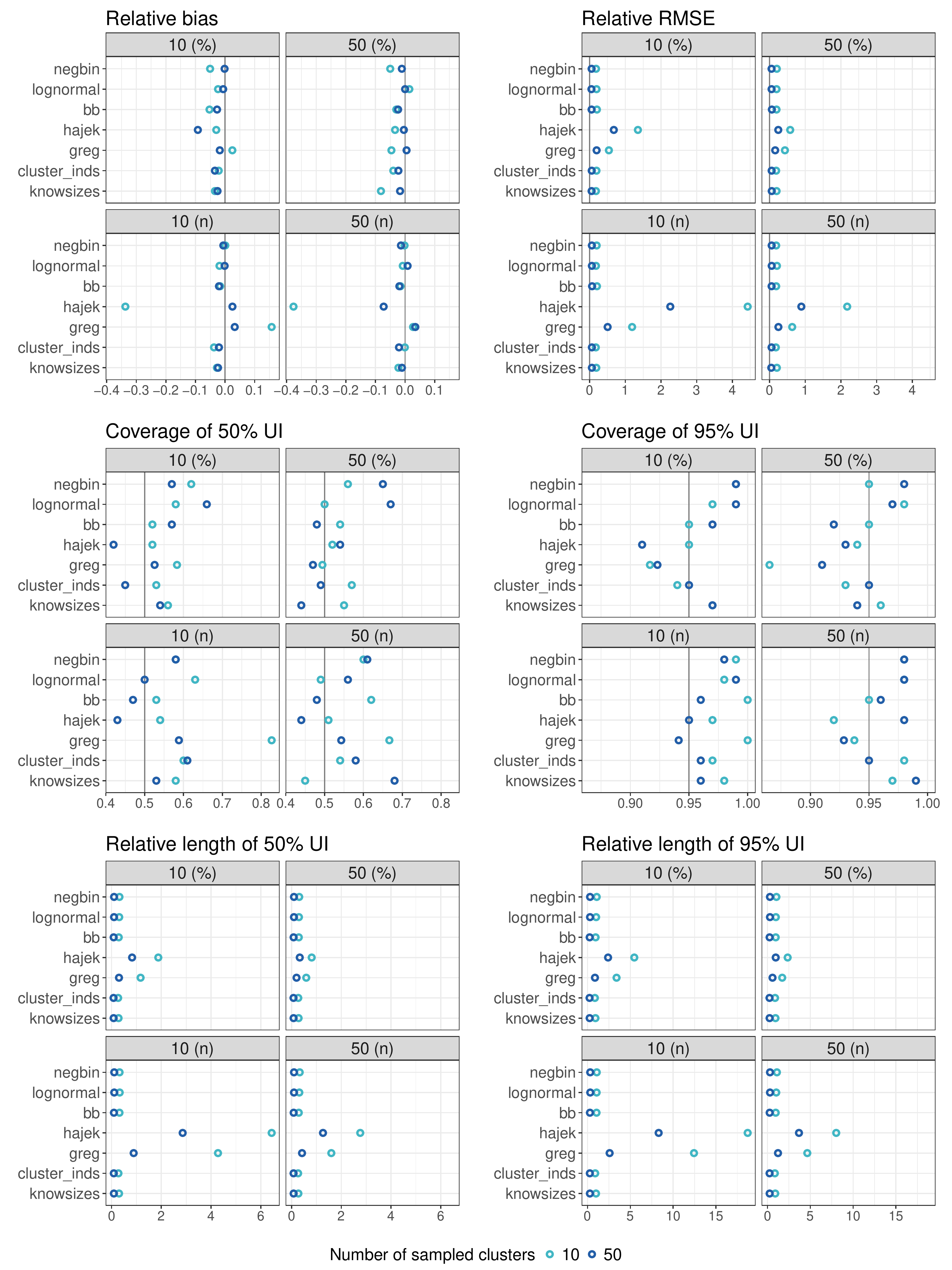}}
	\caption{\em \small Results for continuous $y$ with cluster sizes $N_j$ drawn from a multinomial distribution. The top two plots are for fixed-percentage SRS schemes, and the bottom two are for fixed-number SRS samples. \texttt{negbin}: negative binomial distribution; \texttt{lognormal}: lognormal distribution; \texttt{bb}: Bayesian bootstrap; \texttt{H\'ajek}: the H\'{a}jek estimator; \texttt{greg}: generalized regression estimator; \texttt{cluster\_inds}: the model with random cluster effects but without the cluster size predictor; \texttt{knowsizes}: the model with known population cluster sizes.}
	\label{fig:ybar_cont_mn}
\end{figure}

\begin{figure}
	\centering
	\centerline{\includegraphics[scale=0.5]{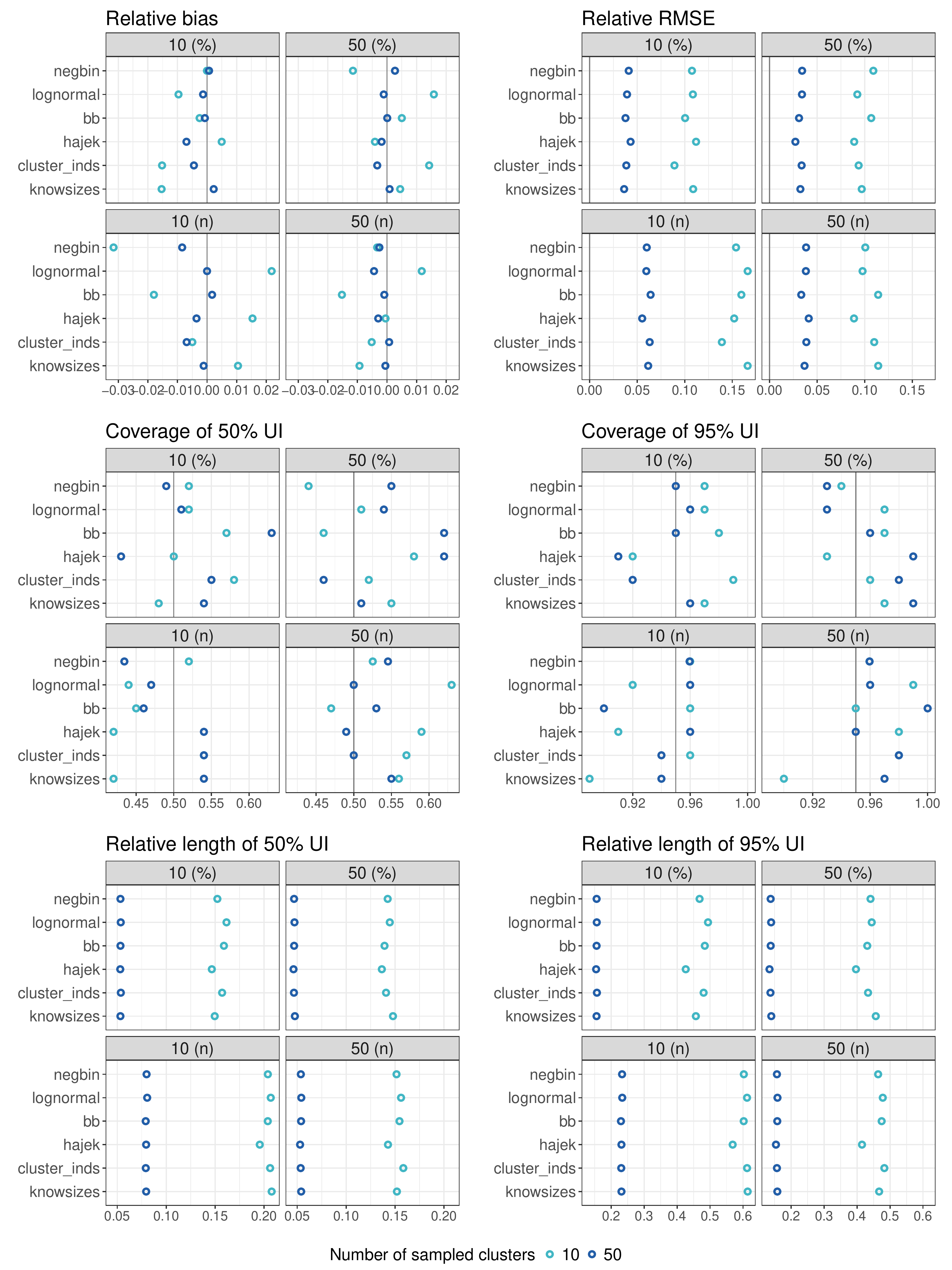}}
	\caption{\em \small Results for binary $y$ with cluster sizes $N_j$ drawn from a Poisson distribution. The top two plots are for the fixed-percentage SRS schemes, and the bottom two represent the fixed-number SRS samples. \texttt{negbin}: negative binomial distribution; \texttt{lognormal}: lognormal distribution; \texttt{bb}: Bayesian bootstrap; \texttt{H\'ajek}: the H\'{a}jek estimator; \texttt{cluster\_inds}: the model with random cluster effects but without the cluster size predictor; \texttt{knowsizes}: the model with known population cluster sizes.}
	\label{fig:ybar_bin_pois}
\end{figure}

\begin{figure}
	\centering
	\centerline{\includegraphics[scale=0.5]{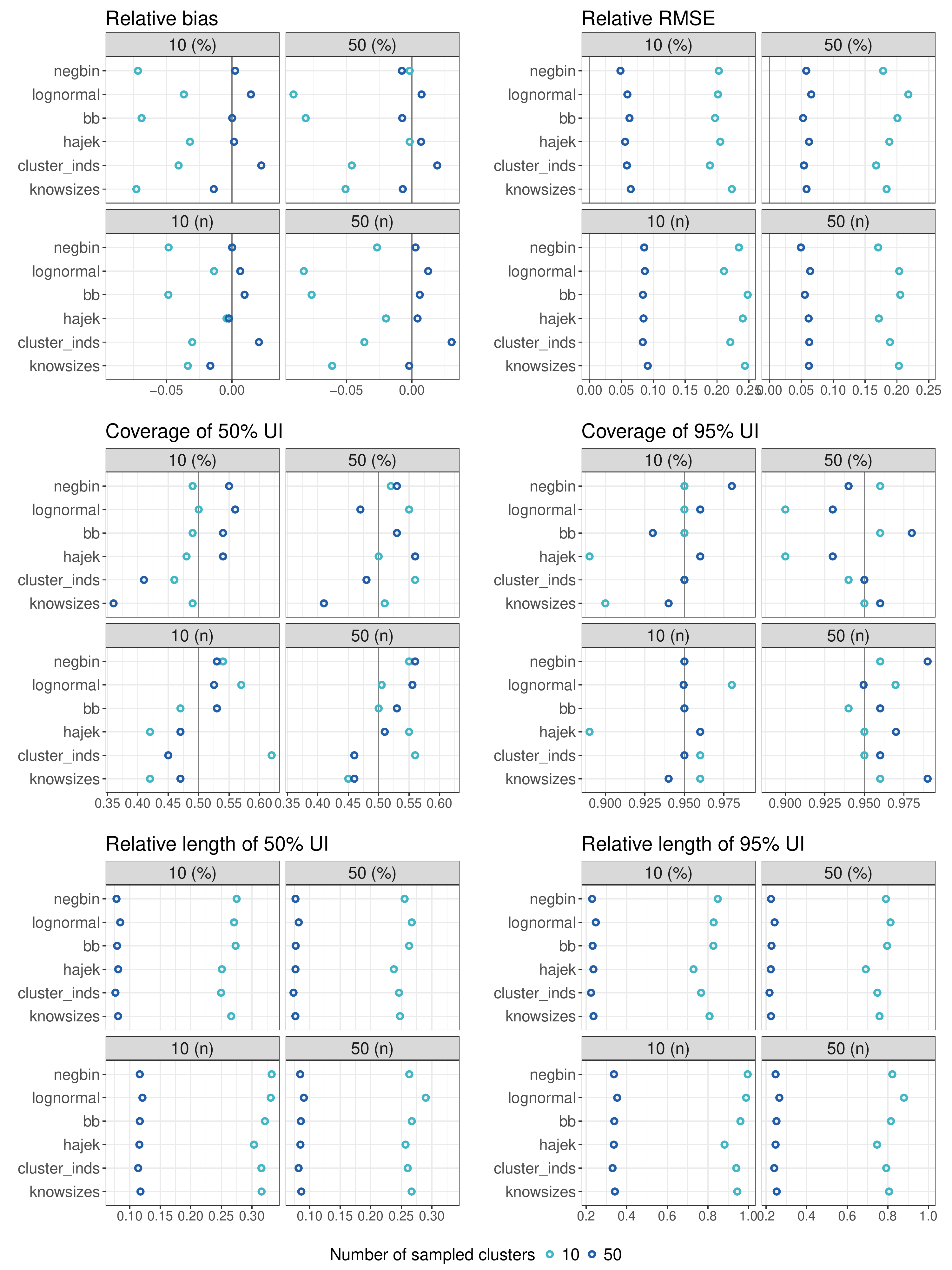}}
	\caption{\em \small Results for binary $y$ with cluster sizes $N_j$ drawn from a multinomial distribution. The top two plots are for fixed-percentage SRS schemes, and the bottom two represent fixed-number SRS samples. \texttt{negbin}: negative binomial distribution; \texttt{lognormal}: lognormal distribution; \texttt{bb}: Bayesian bootstrap; \texttt{H\'ajek}: the H\'{a}jek estimator; \texttt{cluster\_inds}: the model with random cluster effects but without the cluster size predictor; \texttt{knowsizes}: the model with known population cluster sizes.}
	\label{fig:ybar_bin_mn}
\end{figure}

The results of the simulation study are in Figures \ref{fig:ybar_cont_pois} to \ref{fig:ybar_bin_mn}, with each figure displaying a different combination of outcome type (continuous or binary) and population cluster size model (Poisson or multinomial). In each figure, there are six panels displaying the six metrics with which we evaluate the methods: relative bias, relative root mean squared error (RRMSE), coverage of 50\% and 95\% uncertainty intervals, and the average relative widths of the 50\% and 95\% uncertainty intervals. The relative bias is calculated as
$
	\frac{1}{L} \sum_{\ell=1}^L \frac{\overline{y} - \widehat{\overline{y}}_{\ell}}{\overline{y}},
$
where $\overline{y}$ is the true population mean, $\widehat{\overline{y}}_{\ell}$ is the estimated value from the $\ell$-th simulation, and $L$ is the number of simulations. RRMSE is calculated as
$
	\sqrt{\frac{1}{L} \sum_{\ell=1}^L \left(\frac{\overline{y} - \widehat{\overline{y}}_{\ell}}{\overline{y}} \right)^2 }.
$
For the Bayesian methods \texttt{negbin}, \texttt{lognormal}, \texttt{bb}, \texttt{cluster\_inds} and \texttt{knowsizes}, the 50\% (95\%) intervals are calculated from the 25th and 75th (2.5th and 97.5th) percentiles of the posterior predictive distribution for $\overline{y}$. For the classical methods, we rely on asymptotic normal theory and the variance estimators given in Chapter 8 of \cite{sarndal}. The relative widths of the uncertainty intervals are calculated by dividing the width of the uncertainty interval by the true $\overline{y}$ and averaging across the $L$ simulations.

In each plot, the $x$-axis is the metric value and the $y$-axis denotes different models. The panels represent the different within-cluster sampling schemes. The top two plots are for the fixed-percentage schemes, where $n_j = \rho N_j$ for $\rho = 0.1$ and $\rho = 0.5$, $j = 1, \ldots, J_s$. The bottom two plots represent the self-weighting samples, with $n_j = 10$ and $n_j = 50$, $j = 1, \ldots, J_s$. The colors of the circles represent different first-stage sample sizes $J_s$, $J_s \in \{10, 50\}$.

We now describe the results for each of these three comparisons for the four combinations of outcome type (continuous and binary) and population cluster size model (Poisson and multinomial distributions) as explained in the previous section.

Bayesian methods generally yield more efficient inference than classical estimators, particularly with small number of clusters. For continuous $y$, the Bayesian models outperform the design-based estimators, both for the Poisson and the multinomially distributed population cluster sizes in Figures \ref{fig:ybar_cont_pois} and \ref{fig:ybar_cont_mn}, respectively. The differences are rather small when $J_s=50$ but pronounced when $J_s=10$. The H\'ajek estimator has large bias, particularly when the sample is self-weighting, but including auxiliary information as the GREG estimator does greatly reduces the bias. Still, the classical estimators yield unstable results, evident in the high RRMSEs. The Bayesian estimators are preferable here with lower bias and RRMSE, and yield short uncertainty intervals whose coverage rates are close to or above the nominal level. For binary $y$, there is little difference between the Bayesian methods and the H\'ajek estimator when the number of sampled clusters is large, $J_s = 50$. This holds for both the Poisson-distributed cluster sizes in Figure \ref{fig:ybar_bin_pois} and the multinomially distributed cluster sizes in \ref{fig:ybar_bin_mn}. When the number of sampled clusters is small, the H\'ajek estimator and the Bayesian methods has comparable bias and RRMSE. However, the coverage rates for the H\'ajek estimator are often below the nominal level, particularly when the sample is not self-weighting (top row of plots).

Both the parametric and nonparametric approaches are statistically valid and have competitive performances. For continuous $y$ the parametric models \texttt{negbin} and \texttt{lognormal} perform comparably to the nonparametric \texttt{bb} with unbiased estimates and similar RRMSEs in Figures \ref{fig:ybar_cont_pois} and \ref{fig:ybar_cont_mn} particularly under large $J_s$, while coverage is generally higher for the parametric models in Figure \ref{fig:ybar_cont_mn}. For binary $y$, with Poission-distributed cluster sizes the parametric models have a bit higher bias in Figure \ref{fig:ybar_bin_pois}, ranging around 1-1.5\%, while for multinomially distributed cluster sizes in Figure \ref{fig:ybar_bin_mn} the parametric models are less biased than the nonparametric one, especially when the sample is not self-weighting and the number of clusters is small. Coverage rates vary, but are most consistently around or above the nominal level both for the parametric and nonparametric methods. For both continuous and binary $y$, there is little difference in uncertainty interval lengths between the parametric and nonparametric methods.

Incorrectly omitting cluster sizes as cluster-level predictors---that is, using \texttt{cluster\_inds} instead of \texttt{knowsizes}---has little impact when $y$ is continuous for either the Poisson or the multinomially distributed population cluster sizes. The bias, RRMSE, and coverage rates for the two methods are similar in both Figures \ref{fig:ybar_cont_pois} and \ref{fig:ybar_cont_mn}. The differences between \texttt{cluster\_inds} and \texttt{knowsizes} are minor for binary $y$ as well; \texttt{cluster\_inds} does not perform appreciably worse than \texttt{knowsizes} in either Figure \ref{fig:ybar_bin_pois} or \ref{fig:ybar_bin_mn}, the Poisson or the multinomially distributed population cluster sizes. If $y$ and $N_j$ are unrelated, it is not necessary to include $N_j$ in the model, even under PPS sampling; allowing the regression parameters to vary by cluster as in \texttt{cluster\_inds} is sufficient for valid inference. In the application study of Section~\ref{sec:application}, we find that including the cluster sizes as cluster-level predictors will substantially reduce bias and RRMSE with continuous outcome but had subtle difference under binary outcome comparing to the approach only including cluster indicators as random effects models.  It's pivotal to account for the two-stage structure comparing to the PPS design. This shows when the sampling design is complex, including two-stage sampling, cluster sampling, PPS and SRS, some design feature could play a bigger role than others. We recommend controlling for all the design features if possible.

\section{Fragile Families Study Application}
\label{sec:application}

To evaluate the performance of our method in a more realistic survey context, we use a modified version of the Fragile Families study design in conjunction with a presumed outcome model to implement the finite population inference. We would like to use the Fragile Families sampling frame to illustrate the benefits of Bayesian models accounting for the design features. For convenience, we use the outcome estimation model that is the same as the generation model, which assumption can be released as future extensions. 

The Fragile Families study \citep{ffdesign01} divided the 77 U.S. cities with 1994 populations of 200,000 or greater into nine strata based on their policy environments and labor markets. Eight of the strata were for cities with extreme values in at least one of the three policy dimensions under consideration (labor markets, child support enforcement, and welfare generosity), and the ninth stratum was for cities that had no extreme values. One city was selected with PPS in each of the eight extreme strata, with a target sample size of 325 births in each city. In the last stratum, eight cities were selected via PPS, with a target sample size of 100 births in each. There was an intermediate stage of selecting hospitals, which we ignore for the paper illustration. We use the Fragile Families study's city population of 77 cities in 1994 as the sampling frame and implement two-stage cluster sampling under PPS.

As a simulation, we use the city population (divided by 100 for computational convenience) as both the measure of size $M_j$ and the number of units in the cluster $N_j$, though the ultimate unit of sampling in the study was births and number of births in cities should be accounted for. We exclude the three cities that would be selected with probability one for a total of $J=74$ cities. For each unit in the population, we generate an outcome $y$ according to our model in \eqref{eq:dgp}. While the original Fragile Families sampling design involves nine strata, we combine them into a single stratum. As in the actual study design, we sample 16 cities with probability proportional to the city population. In each sampled city, we sample either 325 or 100 births, depending on whether the city is a large- or small-sample city, as designated in \cite{ffdesign01}, which results in a self-weighting sample.

\begin{figure}
	\centering
	\centerline{\includegraphics[scale=0.5]{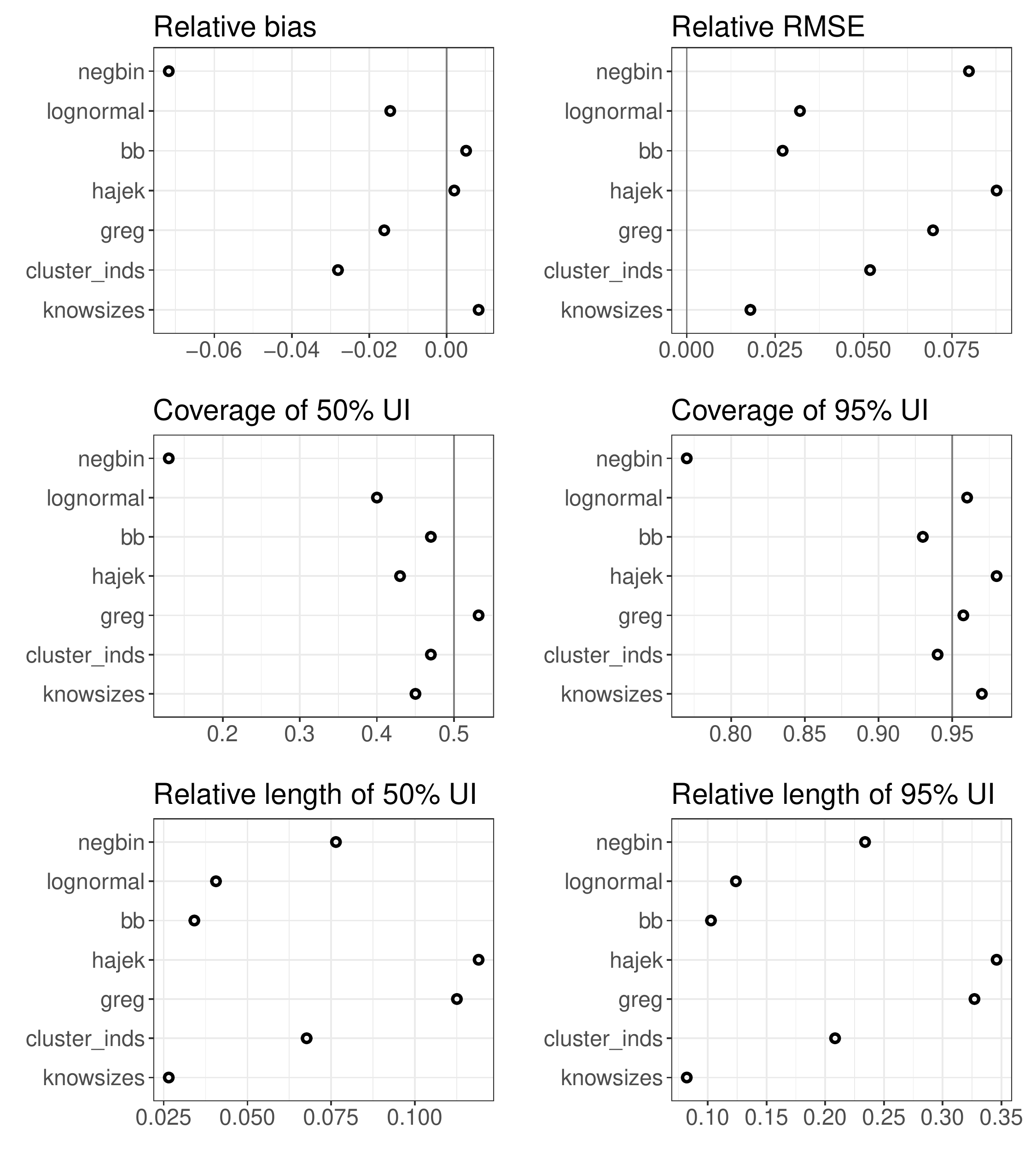}}
	\caption{\em \small Results for continuous $y$ with cluster sizes $N_j$ in the Fragile Families study design. \texttt{negbin}: negative binomial size-biased distribution; \texttt{lognormal}: lognormal size-biased distribution; \texttt{bb}: Bayesian bootstrap; \texttt{H\'ajek}: the H\'{a}jek estimator; \texttt{greg}: generalized regression estimator; \texttt{cluster\_inds}: the model with random cluster effects but without the cluster size predictor; \texttt{knowsizes}: the model with known population cluster sizes.}
	\label{fig:ybar_cont_ff}
\end{figure}

\begin{figure}
	\centering
	\centerline{\includegraphics[scale=0.5]{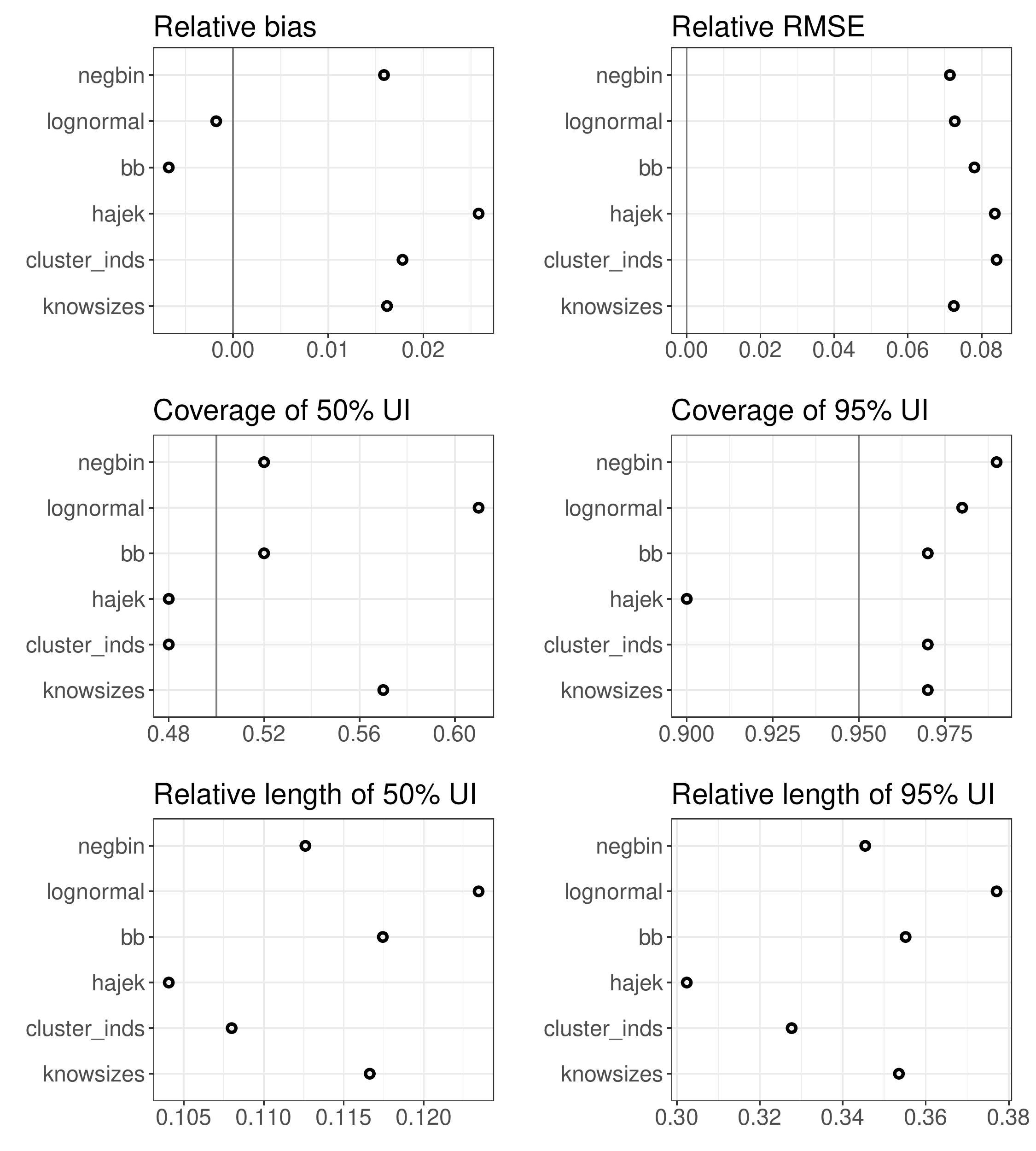}}
	\caption{\em \small Results for binary $y$ with cluster sizes $N_j$ in the Fragile Families study design. \texttt{negbin}: negative binomial size-biased distribution; \texttt{lognormal}: lognormal size-biased distribution; \texttt{bb}: Bayesian bootstrap; \texttt{H\'ajek}: the H\'{a}jek estimator; \texttt{cluster\_inds}: the model with random cluster effects but without the cluster size predictor; \texttt{knowsizes}: the model with known population cluster sizes.}
	\label{fig:ybar_bin_ff}
\end{figure}

Figures \ref{fig:ybar_cont_ff} and \ref{fig:ybar_bin_ff} show the outputs for when the outcome is continuous and binary, respectively, in terms of relative bias, RRMSE, coverage rates and relative widths of 50\% and 95\% uncertainty intervals. The main findings are consistent with the simulation studies.

For continuous $y$ in Figure \ref{fig:ybar_cont_ff}, the Bayesian methods (with the exception of \texttt{negbin}) outperform the design-based estimators in terms of RRMSE and uncertainty interval width and are comparable on bias and coverage. The Bayesian methods yield uncertainty intervals that are less than half the width of those based on the design-based methods, with coverage rate that is close to the nominal level. Among the three Bayesian methods, \texttt{bb} and \texttt{lognormal} perform similarly, and both are better than the \texttt{negbin} assumption. The negative binomial population distribution performs poorly with large bias and RRMSE but low coverage rate. Excluding cluster sizes leads to worse performance, with higher bias and RRMSE and longer uncertainty intervals for \texttt{cluster\_inds} compared to \texttt{knowsizes}.

When $y$ is binary as in Figure \ref{fig:ybar_bin_ff}, we again see that the Bayesian methods yield better results in terms of bias, RRMSE, and coverage than the classical H\'ajek estimator. The uncertainty intervals of the H\'ajek estimator are the shortest but are close to those from the Bayesian methods. Comparing the parametric and nonparametric models, bias and RRMSE are lower with higher coverage rates for \texttt{lognormal} than for \texttt{bb}, though coverage rates for both are above the nominal levels. However, \texttt{lognormal} is conservative since the confidence intervals are longer than those for \texttt{bb}. The \texttt{negbin} has the largest bias but shortest uncertainty intervals with comparable RRMSE and coverage among the three approaches. The effects of excluding the cluster sizes are small, with \texttt{cluster\_inds} having only slightly larger bias and RRMSE than \texttt{knowsizes}.

\begin{figure}
\centering
\includegraphics[scale=.5]{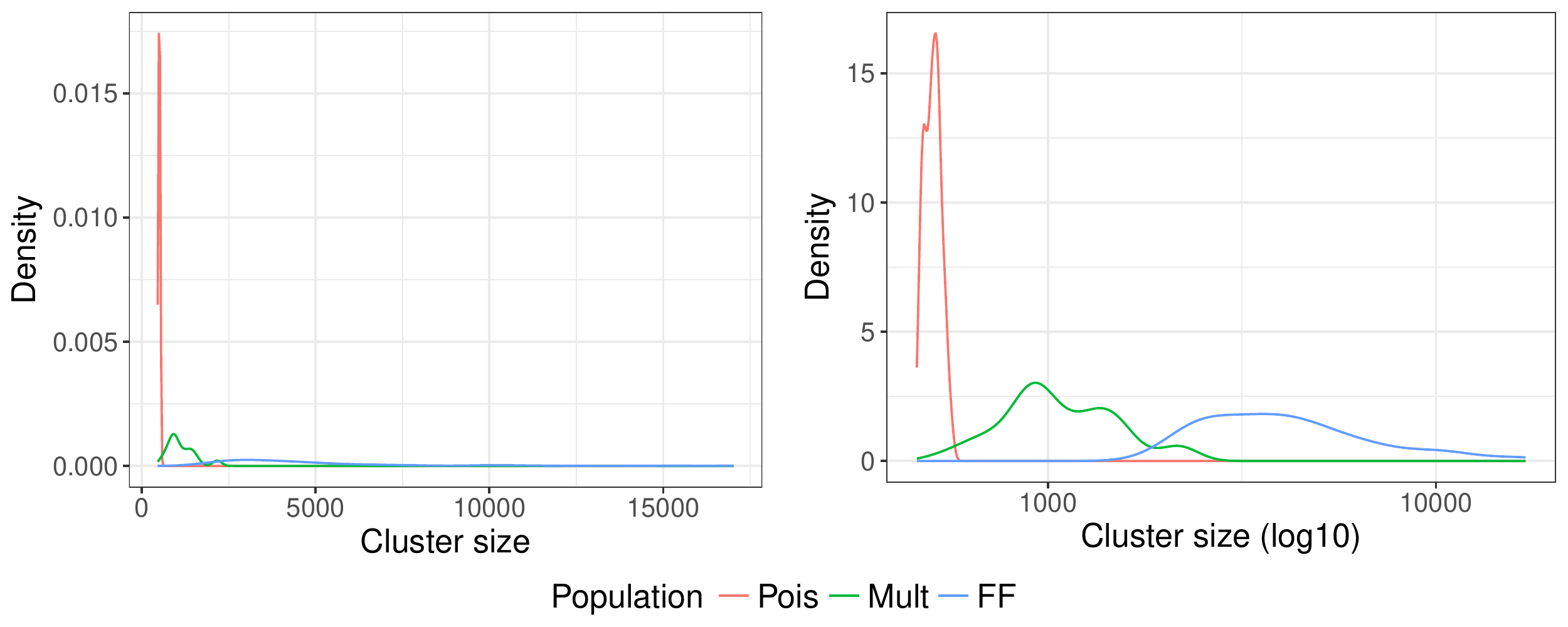}
\caption{\em \small Density plot of 100 cluster sizes drawn from a Poisson distribution with rate 500 (Pois), a Gamma/multinomial distribution (Multi) with a multinomial draw from Gamma(10,1)-distributed samples multiplied by 100, and the Fragile Families (FF) study design. The $x$-axis is on the original scale in the left plot and the log10 scale in the right. This illustrates the distribution of cluster sizes in the Fragile Family study is highly skewed. }
\label{fig:pops}
\end{figure}

To further investigate the population distribution of cluster sizes, Figure \ref{fig:pops} shows the density plots for 100 cluster sizes drawn from the assumed Poisson and multinomial distributions and the 74 (non-certainty for selection) Fragile Families city populations. From the plots, both the Poisson distribution and the multinomial/Gamma distribution used in the simulation study are different from the population distribution of cluster sizes in the Fragile Family study. The cluster sizes in the Fragile Family study are highly skewed. Hence, in the application, the negative binomial size-biased distribution assumption is not appropriate to depict the cluster size population with poor performance. The Bayesian bootstrap method avoids the parametric assumption and yields robust inference, and the lognormal distribution as the size-biased choice is able to capture the skewness and performs competitively, as shown in Figure~\ref{fig:ybar_cont_ff} and Figure~\ref{fig:ybar_bin_ff}. We can modify the parametric assumptions and improve the inference with suitable prior knowledge.

\section{Discussion}
\label{sec:discussion}

We propose an integrated Bayesian model for the finite population inference from a two-stage cluster sample under PPS. Two-stage cluster sampling is popular across health surveys, however, the corresponding model-based inference has methodology challenges. Our method combines predicting measures of size for nonsampled clusters with estimation for the population mean into a single approach that propagates uncertainty from the two steps. We consider both parametric and nonparametric methods for modeling cluster sizes. The parametric models directly account for the unequal probabilities of selection by using the closed-form size-biased version of the underlying population distribution, while the nonparametric Bayesian bootstrap draws from the observed cluster sizes with probabilities that are weighted by the odds of that cluster not being selected. 

While design-based approaches are common in survey inference, variance estimation is often challenging. Current estimation approaches include theoretical approximations~\citep{sarndal} and resampling methods~\citep[e.g.,][]{wolter_2007}. In contrast, our integrated approach yields the posterior distribution for the quantities of interest about the finite population, from which variances, uncertainty intervals, and any other functions can easily be computed. The proposal accounts for the design features in modeling and yields design-consistent inference.

The Bayesian methods generally outperform the design-based estimators and improve inference stability, particularly when the number of sampled clusters is small. The performance of the parametric methods \texttt{negbin} and \texttt{lognormal} is comparable to that of the nonparametric Bayesian bootstrap. When extra information about the population cluster sizes is available, for example, from previous years or similar groups, we can incorporate through the informative prior information. Moreover, the parametric methods are straightforward to implement in Stan, which makes them accessible to researchers whose expertise is in areas outside of statistics or programming. The results for parametric and nonparametric methods are more similar when $J_s = 50$ than when $J_s = 10$ in many of the scenarios our simulation study considered. The parametric method is subject to model misspecification especially under small sample. We recommend using the parametric methods as an initial step and perform model diagnostics. An important diagnostic measure is to check whether the population cluster sizes are highly skewed, as in the case of the Fragile Families setup shown in Figure \ref{fig:pops}. Thus, reasonable prior knowledge of the population distribution of cluster sizes should guide the model choice of parametric or nonparametric approach.

In our study, under binary $y$, the Bayesian methods were less clearly superior to classical methods in estimating the finite population proportion. One possible reason is that few auxiliary or predictive variables are included in the model. However, when the cluster sizes are highly skewed, as in the Fragile Families case, Bayesian methods perform significantly better, in terms of lower bias and more reasonable coverage, than the classical estimators.

There are several interesting directions in which the current research could be extended. First, our simulation has not considered the case where $M_j \neq N_j$ in depth. The natural next step would be to extend the Fragile Families simulation to include the case where the measure of size $M_j$ is the city population, but the cluster size $N_j$ itself is the total number of births in the city. In doing so, we must make some additional assumptions. So far, we have assumed that we know $M_j$ only for the sampled clusters, but what about $N_j$? If both $M_j$ and $N_j$ are only available for sampled clusters, we shall predict both $M_j$ and $N_j$ for the entire population. One idea is to assume that $N_j$ is a function of $M_j$ and use regression models to predict $N_j$ given $M_j$, perhaps the on the log scale to avoid predicting negative cluster sizes and difficulties with cluster sizes ranging over several orders of magnitude. In the Fragile Families study, the correlation between the log of city population $M_j$ and log of total births $N_j$ is $0.78$, so this seems like a promising strategy.

Second, the outcome model can be extended with flexible modeling strategies. To focus on evaluation of different approaches to accounting for the design effect and predicting the nonsampled cluster sizes, for the outcome model, the estimation model we use is the same as the data generation model. In practice, we recommend outcome modeling that is robust against misspecification. Flexible models in the literature can be explored, such as heteroscedasticity assumption, penalized spline regression models, and nonparametric Bayesian models. The multilevel models stabilize estimation via smoothing across clusters. The partial pooling effect can be strengthened with generalized covariance structure, e.g., covariance kernel functions in Gaussian process regression models.

Another direction would be to consider a stratified PPS design as in the original Fragile Families study design. This extension introduces another challenge in that we would need to adjust for the strata structure in our model. For the parametric cluster size models, we would need to partially pool the size parameters (e.g., $\mu, \phi$ in the negative binomial model, $\mu, \sigma$ in the lognormal) across strata, adding another layer of complexity to the model.

Bayesian approaches are well-equipped to account for the design features in the survey data under complex sampling design through hierarchical modeling. Computational software development, such as the use of Stan, makes modeling approaches enhance the advantage. More methodology developments are necessary to incorporate additional information about the sampling into modeling, such as known population size, paradata and auxiliary variables. 

\section*{Acknowledgments}
The work is supported by National Science Foundation research grant awards SES 1534400 and 1534414.

\bibliography{library}
\bibliographystyle{plainnat}

\end{document}